\documentclass[final,3p]{elsarticle}

\usepackage{times}
\usepackage{amsmath}
\usepackage{amsfonts}
\usepackage{amssymb}
\usepackage{graphicx}
\usepackage{xcolor}
\usepackage{microtype}

\journal{Acta Materialia}

\begin{document}

\begin{frontmatter}

\title{The Origin of Deformation Induced Topological Anisotropy in Silica Glass}

\author[fau,lem3,aau]{Sudheer Ganisetti}
\author[fau,max,uds]{Achraf Atila}
\author[lem3]{Julien Gu\'{e}nol\'e}
\author[tubaf]{Aruna Prakash}
\author[hhu]{J\"{u}rgen Horbach}
\author[fsu]{Lothar Wondraczek}
\author[fau,max]{Erik Bitzek\corref{cor1}}

\cortext[cor1]{e.bitzek@mpie.de}

\affiliation[fau]{organization={Department of Materials Science and Engineering, Institute I,Friedrich-Alexander-Universität Erlangen-Nürnberg (FAU)},
	city={Erlangen},
	postcode={91058},
	country={Germany}}
 
\affiliation[lem3]{organization={Université de Lorraine, CNRS, Arts et Métiers, LEM3},
	city={Metz},
	postcode={57070},
	country={France}}
  
\affiliation[aau]{organization={Department of Chemistry and Bioscience, Aalborg University},
	city={Aalborg},
	postcode={9220},
	country={Denmark}}

\affiliation[max]{organization={Computational Materials Design, Max-Planck-Institut für Eisenforschung},
	city={Düsseldorf},
	postcode={40237},
	country={Germany}}

\affiliation[uds]{organization={Department of Material Science and Engineering, Saarland University},
	city={Saarbrücken},
	postcode={66123},
	country={Germany}}
 
\affiliation[tubaf]{organization={Micromechanical Materials Modelling (MiMM), Institute of Mechanics and Fluid Dynamics, Technische Universität Bergakademie Freiberg (TUBAF)},
	city={Freiberg},
	postcode={09599},
	country={Germany}}

\affiliation[hhu]{organization={Institut für Theoretische Physik II, Heinrich-Heine-Universität Düsseldorf},
	city={Düsseldorf},
	postcode={40225},
	country={Germany}}

\affiliation[fsu]{organization={Otto Schott Institute of Materials Research, Friedrich Schiller University Jena},
	city={Jena},
	postcode={07743},
	country={Germany}}

\begin{abstract}
Oxide glasses with a network structure are omnipresent in daily life. Often, they are regarded as isotropic materials; however, structural anisotropy can be induced through processing in mechanical fields and leads to unique materials properties. Unfortunately, due to the lack of local, atomic-scale analysis methods, the microscopic mechanisms leading to anisotropy remained elusive.  Using novel analysis methods on glasses generated by molecular dynamics simulations, this paper provides a microscopic understanding of topological anisotropy in silica (SiO$_2$) glass under mechanical loads. The anisotropy observed in silica glass originates from a preferred orientation of SiO$_4$ tetrahedra at both short- and medium-range levels that can be controlled via the mode of mechanical loading. The findings elucidate the relation between the deformation protocol and the resulting anisotropic structure of the silica network (involving both persistent and transient effects), and thus provide important insight for the design of oxide glasses with tailored materials properties.
\end{abstract}


\end{frontmatter}

\section{Introduction}
Throughout the centuries, oxide glasses have attracted the interest of scientists and engineers due to the sheer number of possible compositions, allowing the development of new glasses that address societal and technological needs~\cite{Shakhgildyan2020}.
Current areas of glass design include flexible screens for mobile devices, optical fibers with super low signal loss, and mechanically resistant solar modules~\cite{Wondraczek2022, Wondraczek2011}. 
Many of these applications result from empirical studies, which have now arrived at a crossroad: for the targeted design of new glasses and to extend further the range of their applications, a fundamental understanding of the relationship between glass composition, processing, structure, and properties is required.
Structurally, oxide glasses consist of glass-forming cations surrounded by oxygen polyhedra.
These polyhedra share corners, establishing a three-dimensional random network. 
The paradigm of a network glass is silica, in which corner-shared SiO$_4$ tetrahedra~\cite{Brckner1970} form Si-O rings of different sizes~\cite{Greaves2007}. 
Due to this amorphous structure, oxide glasses such as silica are typically regarded as being isotropic. 
However, glasses can exhibit anisotropic properties when processed accordingly. 
This is demonstrated most prominently by optical birefringence, which is used to visualize residual or transient mechanical stress embedded in a glass object, for example, monitoring stress fields in Prince Rupert's drops~\cite{Rodrigues2022}.
As another example, homogeneous anisotropy can be induced in glasses by fiber drawing~\cite{Inaba2014, Endo2015, Yang2017, Lund2010}. Such anisotropy, caused by deformation of the viscous melt~\cite{Rountree2009, Sato2013}, is only reversible by thermal annealing. Therefore, in the following, it is referred to as \emph{persistent} anisotropy.
In contrast, glasses can also show elastic or \emph{transient} anisotropy, which is present only during mechanical loading, e.g., as observed in stress-induced reversible birefringence~\cite{Min2008}. 
Mechanical loading can, of course, also lead to persistent anisotropy that remains after switching off the load and  is then reflected, e.g., in the occurrence of residual stresses~\cite{Ballauff2013}. 

Transient anisotropy is of fundamental importance during the deformation and fracture of oxide glasses; it constitutes the initial response of the glass network to mechanical load and thereby defines all subsequent mechanical reactions. The specific directional preference of molecular-scale deformation routes is a direct result of the combination of highly directional bonding among the network-forming species, structural disorder, the resulting occurrence of heterogeneous rigidity~\cite{Pan2021, Benzine2018}, and (if present) further interatomic interactions with reduced directionality (e.g., ionic interactions from cation constituents with lower bond covalency, the so-called network modifiers). However, the specific nature, molecular mechanism, and similarities to persistent anisotropy (which can be studied ex-situ) have been elusive, in particular, for the archetypal case of vitreous silica. Silica is the backbone of practically all contemporary glass materials which exist for widespread application. From a viewpoint of mechanical response, the other chemical components which make up a real-world glass are either highly mobile (such as the network modifiers with predominantly ionic bonding and/or high ion polarizability) or embedded as part of the backbone network (such as secondary network forming species, e.g., B$_2$O$_3$ or Al$_2$O$_3$)~\cite{Lee2005}. Therefore, understanding the transient anisotropy of the silica backbone appears to hold a key towards designing glasses with adapted mechanical performance. 

Persistent anisotropy can be caused by inhomogeneities in the glass that are either becoming aligned or deformed through the deformation process; this effect was denoted \emph{shape anisotropy} ~\cite{Brueckner1996AnisotropicGA}.  
Persistent anisotropy has, however, also been observed in the absence of inhomogeneities.
Early observations of optical anisotropy were explained by frozen-in elastic deformation, assuming that the glass structure consists of elastic and fluid units~\cite{Filon1923}.
Takamori and Tomozawa~\cite{TAKAMORI1977} provided the first systematic review on anomalous birefringence occurring in a range of oxide glasses, which they related to frozen-in strain and microstructural orientation.
The structural origin of mechanically-induced anisotropy in metallic glasses subjected to uniaxial tensile stress during creep was studied using X-ray diffraction by Suzuki et al.~\cite{Suzuki1987}. 
From their observations, they conjectured a model of bond orientation anisotropy, in which 
atoms in the unloaded glass have more nearest neighbor atoms in the direction perpendicular than parallel to the tensile axis~\cite{Suzuki1987}. 

In contrast to metallic glasses, anisotropy in silica glass was shown to be related to the changes at the medium-range structure~\cite{Sato2013, Champagnon2014}. 
Studies on uniaxially compressed silica glass at room temperature (RT) that was unloaded showed a remaining strain~\cite{Sato2013}. This was characterized from the deviation of the first sharp peak from a perfectly radial diffractogram, and it was suggested that the anisotropy was caused by directionality in permanent densification through changes in the medium-range structure~\cite{Sato2013}. 
Similar conclusions were also drawn from Brillouin scattering experiments on uniaxially compressed silica glass~\cite{Champagnon2014}.
In situ mechanical quenching experiments on nanoscale silica spheres were recently performed in a transmission electron microscope~\cite{MACKOVIC2016}. Changes in the Young's modulus after the spheres were uniaxially compressed under the electron-beam were attributed to the reorientation of the Si--O--Si bonds away from the compression axis, leading to an anisotropic structure~\cite{MACKOVIC2016}. 

In contrast to experiments, atomistic simulations allow for a direct, local analysis of the glass structure~\cite{Du2019MD}. 
Molecular dynamics (MD) simulations showed that mechanical loading could induce structural anisotropy~\cite{Ghemid2007, Rountree2009, Bidault2016}. 
In particular, a change of the medium-range structure was demonstrated using the fabric tensor to characterize the mean orientation of the Si-O-Si bonds. It was shown that silica glass plastically deformed under shear can exhibit structural anisotropy even in the unloaded state~\cite{Rountree2009}. 
An MD study that tried to mimic the drawing process of silica fibers at high temperatures using uniaxial tensile tests in a fully periodic system showed that the normals of small rings tend to orient parallel to the stress axis during the drawing and cooling processes~\cite{Bidault2016}. 
In contrast, the 6-membered rings have their surface normals oriented orthogonal to the loading axis~\cite{Bidault2016}.  
These preferential orientations persist when cooled to room temperature (under load), leading to an anisotropic glass structure and anisotropic elastic properties~\cite{Bidault2016}. 
For a silica glass subjected to c\texttt{}ompression at 1000 K, it was observed that the three-membered rings have their plane normal orthogonal to the compression axis~\cite{Ghemid2007}.

While MD simulations have provided some insights into the experimental findings, the use of different flow regimes and/or loading modes has hindered a one-to-one comparison between MD studies and experiments, which led to the lack of clear conclusions on the origins of the structural anisotropy observed in silica glass.
In particular, it is not clear whether mechanically induced persistent anisotropy implies similar structural changes than transient anisotropy and whether persistent anisotropy caused by deformation at room temperature
is comparable to anisotropy produced around or above the glass transition temperature ($T_g$), e.g., during wire drawing.

Here we show that transient and persistent anisotropy induced by uniaxial deformations at RT have clearly different structural origins. 
Using stereographic projections, we demonstrate that during uniaxial tension and compression, the SiO$_4$ tetrahedra and the Si--O rings reorient with respect to the loading axis, leading to transient anisotropy. 
Persistent anisotropy in the unloaded glass is caused by permanent changes in bond topology and is more pronounced after compression than after tension due to the larger plastic strains reached in the former.
In addition, we demonstrate that persistent anisotropy induced through loading at temperatures around $T_g$ is structurally different from anisotropy produced by RT deformation.

\section{Materials and Methods}
\subsection{Glass preparation}
The atomic interactions were modelled by the first principles based polarizable force field developed by Kermode et al.~\cite{Kermode2010}. 
Additional simulations using the BKS potential~\cite{vanBeest1990, Rajappa2014} showed no qualitative differences; see supplementary material Fig.~S1.
The initial, pristine glass sample was prepared by deforming and rescaling an $\alpha-$quartz structure containing 72000 atoms into a cubic simulation box to match with the experimental density of 2.2 g cm$^{-3}$ of silica glass~\cite{Mozzi1969}. Periodic boundary conditions (PBC) were used along all directions.
This sample was molten at 5000 K in the NVE ensemble for 10 ps, kept at this temperature for 10 ps in the NVT ensemble, and 100 ps in the NPT ensemble at 0 MPa while maintaining a cubic simulation box. 
The sample was cooled to 300 K in the NPT ensemble at a cooling rate of 10$^{13}$ K s$^{-1}$ and 0 MPa,
resulting in a glass with a density of 2.39 g cm$^{-3}$, see also Tab.~S1 in the supplementary material.
For this quenching rate, the glass transition temperature was determined to $T_g=3000$K.
Additional glasses were prepared with cooling rates of $2\times10^{13}$ K s$^{-1}$ and $1\times10^{14}$ K s$^{-1}$.
Nos\'e-Hoover thermo- and barostats were used to perform the constant temperature and pressure simulations~\cite{Hoover1985, Parrinello1981}. 
A timestep of 0.1 fs was used for the MD simulations at temperatures above 1000 K and 1 fs for temperatures below.

\subsection{Deformation}
The equilibrated stress-free isotropic glass sample ('pristine sample') was subjected to uniaxial deformation by tension or compression at 300 K.
The deformation was applied by homogeneously scaling the samples along the $x$-direction at a strain rate of  5$\times10^{8}$ s$^{-1}$, while maintaining zero stress in the orthogonal directions.
Additional deformations were performed with strain rates of  1$\times10^9$ s$^{-1}$ and 5$\times10^9$ s$^{-1}$, see supplementary material Fig.~S2.
At strains of $\epsilon_{x}$ (where $x$ = 2\%, 5\%, 7\% and 17\%, during tension and $x$ = 2\%, 5\%, 7\%, 17\%, 20\%, 25\%, 30\%, 35\% and 40\%, during compression) the sample was unloaded back to the stress-free state.
Additional simulations to investigate the effect of the cooling rate and deformation temperature on the stress-strain response and anisotropy index are shown in Fig.~S3 and Fig.~S4 in supplementary materials. 

Deformations at higher temperature to mimic the wire-drawing process were performed at $T_g$. 
Here, the melt was first cooled under temperature and pressure control from 5000 K to 3000 K with a cooling rate of 10$^{13}$ K s$^{-1}$ while keeping zero stress in all axes. 
The stress-free sample at 3000 K was subjected to uni-axial tension or compression up to 17\% strain along the $x$-axis while keeping zero stress on the other two orthogonal axes. After the deformation, the samples were quenched to 300 K while allowing the box and atoms to relax the accumulated stress using a cooling rate of 10$^{14}$ K s$^{-1}$.
The resulting glasses have densities of 2.35 g cm$^{-3}$ for the glasses pre-deformed in tension and 2.43 g cm$^{-3}$ for the one pre-deformed in compression at $T_g$.

\subsection{Anisotropy index}
The change of the orientation of a pair of atoms during the loading and unloading is monitored and quantified by the anisotropy index ($\alpha$). The anisotropy index is a scalar value obtained by calculating the eigenvalues ($\lambda_i$) of the fabric tensor $F = <n\otimes n>$~\cite{Rountree2009} as given in eq. \ref{anisoindex}, and is normalized in a way that it will vary between 0 and 1,
\begin{equation}
	\label{anisoindex}
	\alpha = \sqrt{\frac{3}{2} \sum_{i=1}^3 \left(\lambda_i -\frac{1}{3}\right)^2 }
\end{equation}
For an isotropic material, $\alpha$ = 0, and in the case of full anisotropy $\alpha$ = 1, higher values mean higher anisotropy in the medium. We computed $\alpha$ for Si--O, Si--Si, and O--O pairs, and it is referred to as $\alpha(Si-O)$, $\alpha(Si-Si)$, and $\alpha(O-O)$, respectively.

\subsection{Anisotropic radial distribution function} 
To capture the short-range order's directionality, a suitable expansion of the RDF onto spherical harmonics must be utilized. Particularly, we take the following expansion into spherical harmonics Y$_{lm}(\theta, \phi)$
\begin{equation}
	g(r) = \sum_{l=0}^{\infty} \sum_{m=-l}^{l} g_{lm}(r)Y_{lm}(\theta, \phi)
\end{equation}
where $l$ and $m$ are the degree and order of the spherical harmonic and $g_{lm}$ is the expansion coefficient.
During a uniaxial deformation, the deformation axis will be the symmetry axis. 
Checking the spherical harmonic functions, $Y_{20}$  best captured this behavior. 
The expansion coefficients of the RDF into the $Y_{20}$ spherical harmonic is given as,
\begin{equation}
	g_{20}^{\alpha\beta} = \frac{\sqrt{15}V}{\sqrt{16\pi}N_{\alpha}N_{\beta}}\left< \sum_{i}^{N_{\alpha}}\sum_{j\neq i}^{N_{\beta}} \delta (|r_{i}^{\alpha}-r_{j}^{\beta}|-r)\frac{2(z_{i}^{\alpha} - z_{j}^{\beta})^{2} - ((x_{i}^{\alpha} - x_{j}^{\beta})^{2} + (y_{i}^{\alpha} - y_{j}^{\beta})^{2})}{(r_{i}^{\alpha} - r_{j}^{\beta})^{4}}\right>,
\end{equation}
where $\alpha$ and $\beta$ are the atom types, $N_{\alpha}$ and N$_{\beta}$ are their corresponding number of atoms, $V$ is the volume of the simulation box, $x$, $y$, and $z$ are the atomic position in Cartesian coordinates.  
This function is positive when more bonds between atom types $\alpha$ and $\beta$ of length $\langle|r_i -r_j|\rangle$ are oriented in parallel to the loading axis than orthogonal to it, and negative when there are more bonds of a certain length orthogonal than parallel to the loading axis.

\subsection{Stereographic projections of tetrahedra face normals and rings}
To the best of our knowledge, the stereographic projection is used for the first time to characterize the structure of oxide glasses in this work. 
The steps that we used to adapt this method to our needs are detailed in the supplementary materials. First, the tetrahedral face normal calculation was made by finding the nearest neighbors of each Si atom. Then, the four-fold coordinated Si atoms were identified, each Si atom and its four nearest neighboring O atoms were separated, these five atoms together were referred to as tetrahedron, each tetrahedron has four faces, and each face has three atoms. Next, the normal vector of each face of the tetrahedra was computed using the position vectors of the three O atoms. Lastly, these vectors were normalized and then stereographically projected on the required plane as described by the procedure detailed in the supplementary materials.

For the ring normals, a similar approach was used where all rings of a given sample were identified using the RINGS code~\cite{LeRoux2010}. Each ring's Si and O atoms were separated; see Fig.~S5(d) for an example with a six-membered ring. The center of mass of each ring was computed based on the position of the atoms, see Fig.~S5(e). Then, each ring was discretized into triangles with 2 neighboring Si atoms and the center of mass of the corresponding ring. After that, the normal vector of each triangle was computed, see Fig.~S5(f). 
The ring normal is determined as the normalized average of all its triangle normals, see Fig.~S5(g). Lastly, all the normals of the same-sized rings were projected on the required plane using stereographic projection by following the procedure above (see Fig.~S5(h - j)).

\subsection{Bond statistics}
Bond statistics following~\cite{Luo2016} was used to determine the number of atoms with new, switched, or broken bonds. This was achieved by comparing the deformed configuration with the pristine glass configuration and finding the differences in bonding topology between them. Each atom has a unique ID. 
This allows to determine atoms with broken or added bonds by comparing the number of their nearest neighbors with respect to the reference configuration. 
Atoms with switched bonds have the same number of nearest neighbors as in the reference configuration but have at least one nearest neighbor with a different ID than in the reference configuration.

\section{Results and Discussions}
\subsection{Deformation of silica  glass and characterization of structural anisotropy}
\begin{figure}[ht]
\centering
\includegraphics[width=0.7\textwidth]{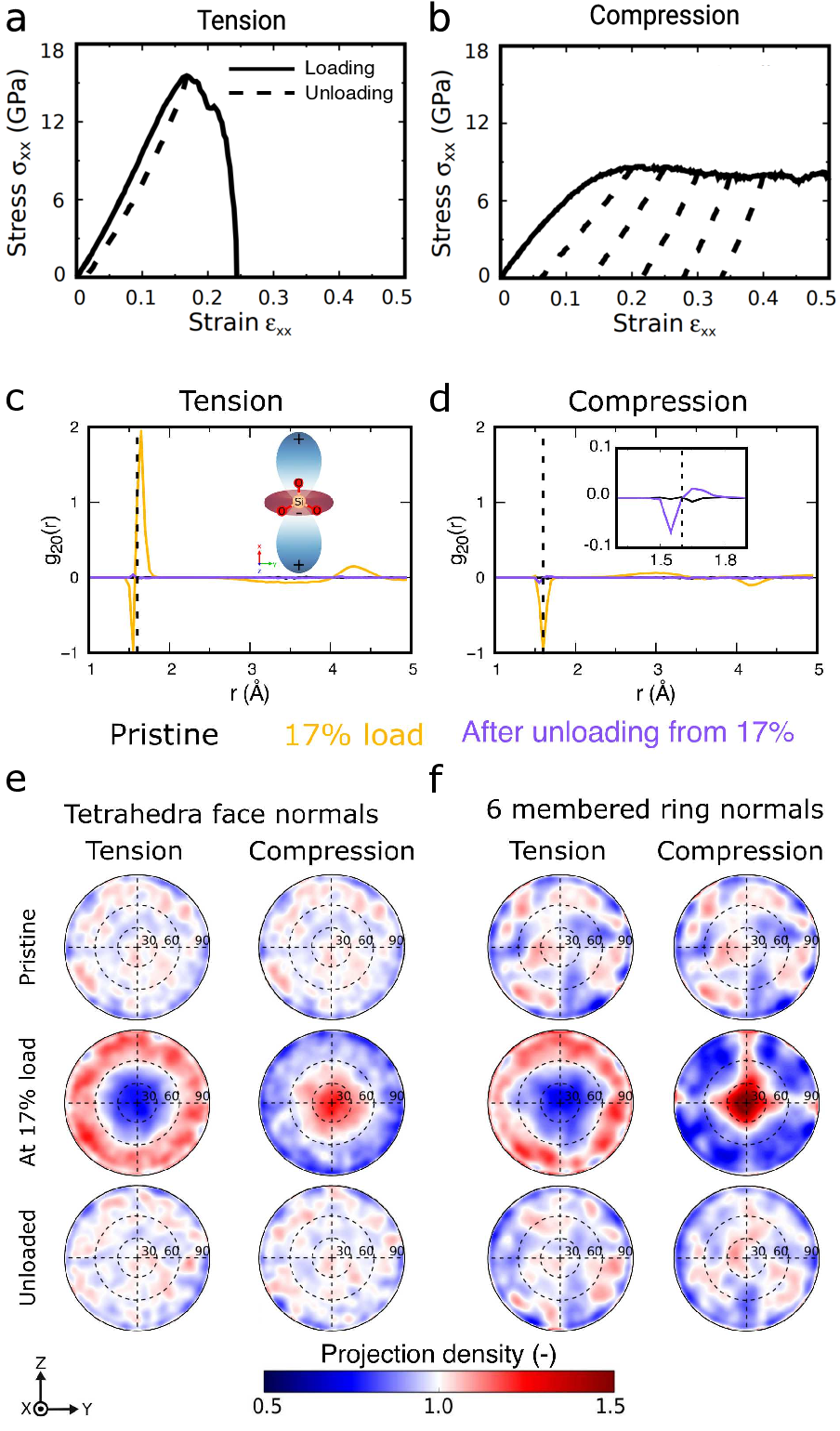}
\caption{\label{Stressanisofact}\footnotesize Top row: Stress-strain response in  tension (a) and compression (b) ($\dot{\epsilon}$ = 5$\times10^{8}$ s$^{-1}$, $T = 300$ K). 
Middle row: Projection of the Si-O pair distribution function o\texttt{}nto the spherical harmonic $Y_{20}$ for glasses at different stages of deformations in tension (c) and compression (d).
Solid black, yellow, and purple lines indicate the samples at initially 0\% strain, loaded to 17\% strain, and stress-free, unloaded from 17\%\ strain. 
The dashed lines indicate the position of the average Si--O bond length in the pristine sample. 
If there are more bonds of the same length $r$ in the loading $(x)$ direction than in the orthogonal directions, $g_{20}(r)$ is positive and vice-versa for negative $g_{20}(r)$.
Bottom row: Stereographic projection maps of the tetrahedra face normals (e) and 6-membered rings normals (f) at different stages of deformation. 
The color indicates the density in the normalized projection map, see Fig:~S5 in the supplementary material for the definition of stereographic projections.
}
\end{figure}

The usual structural characterization of the simulated silica glass at RT are shown in the supplementary material, Fig.~S6, and these are consistent with the data reported in the literature from both experiments~\cite{Greaves2007} and simulations~\cite{Kermode2010}.
The stress-strain curves of pristine silica glass subjected to tension and compression at RT are shown in Fig.~\ref{Stressanisofact} and show brittle fracture in the case of tension and plastic flow in compression. 
Unloading simulations to zero stress were performed starting from different values of the maximum strain as depicted in Fig.~\ref{Stressanisofact}(a and b). 
Unloading from strains beyond 12\% tensile strain resulted in a permanent plastic strain. 
For compressed samples, a permanent plastic strain was observed for unloading from strains beyond 10\%.
The densities after unloading from different loads are provided in the supplementary material, Tab.~S1.

\clearpage

During straining, the radial distribution function and the bond angle distribution are clearly changed from their pristine state, see Fig.~S6 in the supplementary material. 
However, after unloading, only negligible differences are observed, and these isotropic measures cannot be used 
to characterize anisotropy.
Therefore, following earlier studies~\cite{Rountree2009}, the anisotropy was first quantified by the anisotropy index $\alpha$ (see method section for details), which is shown in Fig.~S7 in the supplementary material.
The increase of $\alpha$(Si-Si) and $\alpha$(Si-O) during the loading up to the yield stress shows that the deformation induces some alignment in the Si--Si and Si--O bonds.
This transient anisotropy during loading was present irrespective of the deformation mode.
Significant persistent anisotropy was only observed in compression, as shown in Fig.~S7(d).
Simulations performed at different strain rates and on samples produced with different cooling rates showed that the observed trends regarding transient and persistent anisotropy do not depend on those parameters.
However, the degree of anisotropy and the remaining plastic strain depends on the cooling and strain rates; see Figures~S3 and S2 in the supplementary material.
Deformation at higher temperatures showed that anisotropy is affected by the deformation temperature (see supplementary material Fig.~S4).

To characterize in detail preferred orientations in short-range order, we use the spherical harmonic expansion of the pair distribution function~\cite{hess2015}. 
In the case of an axial symmetry, such as in the current study, the radial distribution function is projected on the spherical harmonic $Y_{20}$; see the methods section for more detail. 
Figure~\ref{Stressanisofact}(c and d) shows the spherical harmonic coefficient $g_{20}(r)$, which describes the intensity of spherical harmonic term $Y_{20}$ for the pristine glass, loaded to 17\% in tension and compression and unloaded from 17\% strain to zero stress. In the following, we will focus on these samples; loading up to and unloading from higher strains resulted in qualitatively similar behavior.
For the pristine glass, $g_{20}(r)$ shows that the Si--O bonds are distributed randomly without preferential orientation and indicates that pristine silica glass has an isotropic short-range structure. 
During tension, Fig.~\ref{Stressanisofact}(c), more Si--O bonds that are longer than the average bond length in a pristine sample ($\langle l_\text{Si--O}\rangle =1.6$\AA\, see supplementary information, Fig.~S6) are aligned in the loading direction than in directions orthogonal to it, while shorter bonds tend to be oriented orthogonal to the loading axis.
The isotropic structure is mostly recovered when unloaded back to the zero-stress state, as indicated by the absence of noticeable peaks after the unloading. 
In contrast, the behavior during compression is different from the one in tension, and only one negative peak centered around $\langle l_\text{Si--O}\rangle$ is observed, indicating that there are more Si--O bonds of average length oriented perpendicular to the loading axis than parallel to it.
Interestingly, the unloaded sample after compression shows qualitatively the same $g_{20}(r)$ signature as the glass loaded \emph{during tension}: shorter than average bonds oriented mostly orthogonal to the loading axis, and more elongated bonds along the loading axis (Fig.~\ref{Stressanisofact}(d)).

The glass structure can be further characterized by determining the orientation of tetrahedra face normals, see the methods section and the supplementary material for more details on the method. 
Figure~\ref{Stressanisofact}(e) shows the normalized stereographic projections of tetrahedra face normals on the $yz$-plane in both tension and compression for the pristine, loaded, and unloaded samples. 
As shown for the pristine silica glass, the normalized stereographic projection of tetrahedra face normals shows that the tetrahedra are randomly oriented in all possible directions without having any preferential orientation.
In the loaded sample in tension, the normalized stereographic projections indicate that most tetrahedra faces are oriented perpendicular to the tensile axis.
At the same level of strain in compression, the normalized stereographic projection shows that most of the tetrahedra faces are oriented perpendicular to the compression axis. 
In the unloaded samples, the normalized stereographic projections show a uniform orientation of the tetrahedra. 

The ring statistic presented in Fig.~S6 shows that the number of 6-membered rings, which represent the majority of the rings in our glass, is only slightly affected by the deformation.
Figure~\ref{Stressanisofact}(f) shows the normalized stereographic projection of 6-membered ring normals for the pristine silica, loaded, and unloaded samples (see the methods section and supplementary materials Fig.~S5 for the computation of ring normals).
Although a uniform orientation of ring normals is expected in pristine silica, insufficient averaging due to the small sample leads to small deviations of the projected 6-membered ring normals from a uniform density, see Fig.~\ref{Stressanisofact}(f). 
However, a clear orientation of the 6-membered ring plane normals away from the loading axis is visible for the sample loaded in tension.
Under compression, the loaded samples show a strong preferential orientation of 6-membered ring-plane normals along the compression axis. 
After unloading to zero stress, a uniform orientation of 6-membered ring normals is observed for both tension and compression.

\subsection{Evolution of bond topology under load}

\begin{figure}[ht!]
	\centering
	\includegraphics[width=0.8\textwidth]{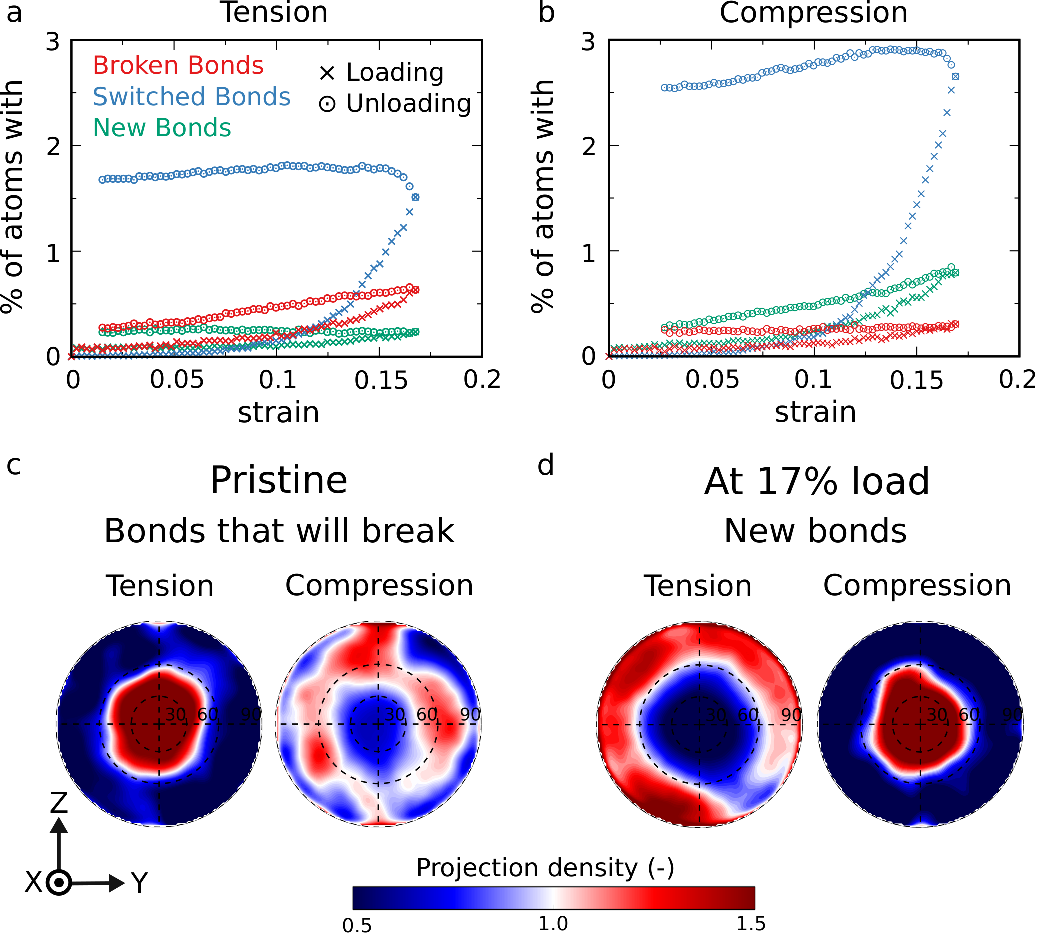}
	\caption{\label{bondstatus}\footnotesize Top: Evolution of broken (red), switched (blue), and new (green) bonds during tension (a) and compression (b).
		Solid lines correspond to loading to 17\% strain, and dashed lines mark the unloading from 17\% strain.  
		Bottom: Stereographic projections for tension and compression showing the orientation of bonds that will be broken during the straining to 17\% (c) and for bond that formed during the straining to 17\% (d).  
		The orientation of the bonds that will be broken is shown in the pristine glass, while that of the newly formed bonds is shown in the loaded samples. 
	}
\end{figure}

Figure~\ref{bondstatus}(a), shows the evolution of new, switched, and broken bonds during tensile loading to 17\% strain and unloading from it. The methodology of computing the bond statistics follows ref.~\cite{Luo2016} and is described in the method section.
The percentage of atoms  which switched neighbors before yielding at around 12\% tensile strain is less than 0.5\%. 
With yielding, this fraction increases to 1.5\% at 17\% strain.
During unloading, some more bond switching occurs, and then the percentage of atoms with switched bonds stays roughly constant, indicating that the new bonding topology is maintained with unloading.
As expected for tensile loading, the percentage of atoms with broken bonds increases, but less strongly than the number of switched bonds, and only a negligible number of atoms adds bonds.
While the number of newly formed bonds stays constant, some broken bonds are healed during unloading.
During compression, Fig.~\ref{bondstatus}(b), only very few bonds are broken,
and more bonds are formed than switched until yield when the percentage of bond switching significantly increases from about 10\% strain on.
However, even in compression, the percentage of atoms with changed bonding topology remain below 3\% at 17\% strain.
After unloading from 17\% strain to zero stress, the percentage of atoms with broken bonds decreases slightly, and the percentage of atoms with new bonds reduces from 0.9\% to 0.3\%, 
while the number of switched bonds further increases during the initial stages of unloading and then only slightly decreases. 

\subsection{Origin of transient and persistent anisotropy}{\label{structural_origin}}

The orientation of the bonds that will be broken at 17\% strain and the bonds that will be newly formed at 17\% strain are shown in Fig.~\ref{bondstatus}(c and d).
Although in tension only relatively few bonds are broken before reaching the maximal tensile stress,
the bonds that break were already in the pristine sample clearly oriented parallel to the loading axis.
In compression, the very few bonds that break were preferably oriented at about 60\textdegree{} to the compression axis before deformation.
Most of the bonds that were newly formed during tension make at 17\%  strain an angle between 60 and 90\textdegree{} with the tensile axis.
The new bonds formed during compression make at 17\% compression preferentially an angle between 0 and 30\textdegree{} with the compression axis.

In the elastic regime in compression, the strain is accumulated by reducing Si--O bond lengths in the direction of compression and/or by changing Si--O--Si angles, leading to transient anisotropy.
The major portion of densification in this regime is associated with rearranged network topology upon reduced Si--O--Si bond angles, which is in consistent with Trease et al. ~\cite{trease2017bond}.
The reduced Si--O bond length is in consistent with Benmore et al. ~\cite{benmore2010structural}.
When the sample is unloaded from this regime, the stored elastic strain is released, and the silica glass returns to its initial state of isotropy. 
Unloading from the plastic regime resulted in a finite plastic strain at zero stress, and the unloaded glass showed structural anisotropy. 
The transition from elastic to plastic regime is correlated with a strong increase in the number of atoms with switched bonds.

Persistent anisotropy can thus be understood in terms of changes in bond topology through plastic deformation and bond switching, as opposed to simple ring deformation reactions.
As shown in Fig.~\ref{bondstatus}(c and d), this bond switching is not random.
Due to the Poisson expansion orthogonal to the compressive axis, bonds in this direction are elongated and, therefore, more prone to bond switching or breaking.
New bonds are then preferentially formed along the compression axis, where atoms are brought closer together. 
Upon unloading, the new bonding topology is maintained (only few bond switching events during unloading, Fig.~\ref{bondstatus}(b)).
However, the newly formed bonds along the loading axis become now elongated as the atoms move further away. 
The initial bonding partner, however, can just relax, leading to shorter bonds orthogonal to the loading axis. 
This explains the $g_{20}(r)$ signature of the persistent anisotropy introduced by pre-compression,
Fig.~\ref{Stressanisofact}(d). 
The same rationale applies to glasses pre-deformed in tension, however, there the significantly reduced tendency for bond-switching compared to compression leads to only weak anisotropy.

Compression of silica glass is well known to result in densification through  (1) decreasing the Si-O-Si bond angles, (2) by enlarging the Si-O bond lengths, and (3) by increasing the number of over-coordinated Si \cite{Trease2017,benmore2010structural}. 
It is, however, important to note that a) densification is usually observed as persistent
densification, i.e., after load removal; and that b) densification is commonly examined after samples were subjected to 
hydrostatic strain states or similar cases, in which the flow of material is restricted to a certain volume like in the 
case of indentation. 
Our observations show that under purely uniaxial compression, i.e. by allowing Poisson expansion, the densification is much less than under comparable hydrostatic strains, see Table S1 in supplementary material.
The densification in our simulations of uniaxial compression is due to the reduction of Si-O-Si bond angles, see Fig.~S6d in the supplementary material and not by bond shortening (Fig.~S6b) or by overcoordinated Si atoms (Fig.~S8).
Changes in bonding topology, bond lengths and bond angles after compression, 
can, however, due to different boundary conditions, very well be different for uniaxially compressed and indented silica glass. 
This is important when comparing our results to complementary work, e.g., by He et al. \cite{He2020,He2021}.
Furthermore, scalar measures like bond-length distributions can not provide significant information on densification by uniaxial compression, as changes in different directions can cancel each other out. 
Finally, the notion of persistent densification implies also the possibility of transient densification which, as our work on anisotropy has shown, might have different topological signatures compared to persistent densification.

While the deformation-induced anisotropy was studied here only for silica glass, the results can be used to discuss what can happen when modifiers (e.g. Na$_2$O) are present in the glassy matrix. 
Silica glass has a high structural homogeneity, which renders it difficult to induce structural anisotropy. 
A higher level of structural anisotropy is expected in glasses that contain structural inhomogeneities due to the presence of highly directional covalent bonds, i.e., Si-O and modifiers-O bonds with less to no bond directionality.
The presence of modifiers within the glass leads to depolymerization of the structure by transforming the bridging oxygens (BO) to non-bridging oxygen (NBO). 
For example, meta-silicate and meta-phosphate glasses have a chain-like structure made of glass former tetrahedra that are linked by BOs. 
These chains are separated by modifier-NBO bonds, which have low bond strength compared to Si-BO or P-BO bonds. 
Breaking these chains under load can be harder than breaking the bonds associated with the non-bridging oxygen. 
As reported by Yang et al. \cite{Yang2017} and Endo et al. \cite{Endo2015}, a higher level of anisotropy can be induced by freezing these chains into a preferential orientation.

\begin{figure}[!ht]
	\centering
	\includegraphics[width=0.7\textwidth]{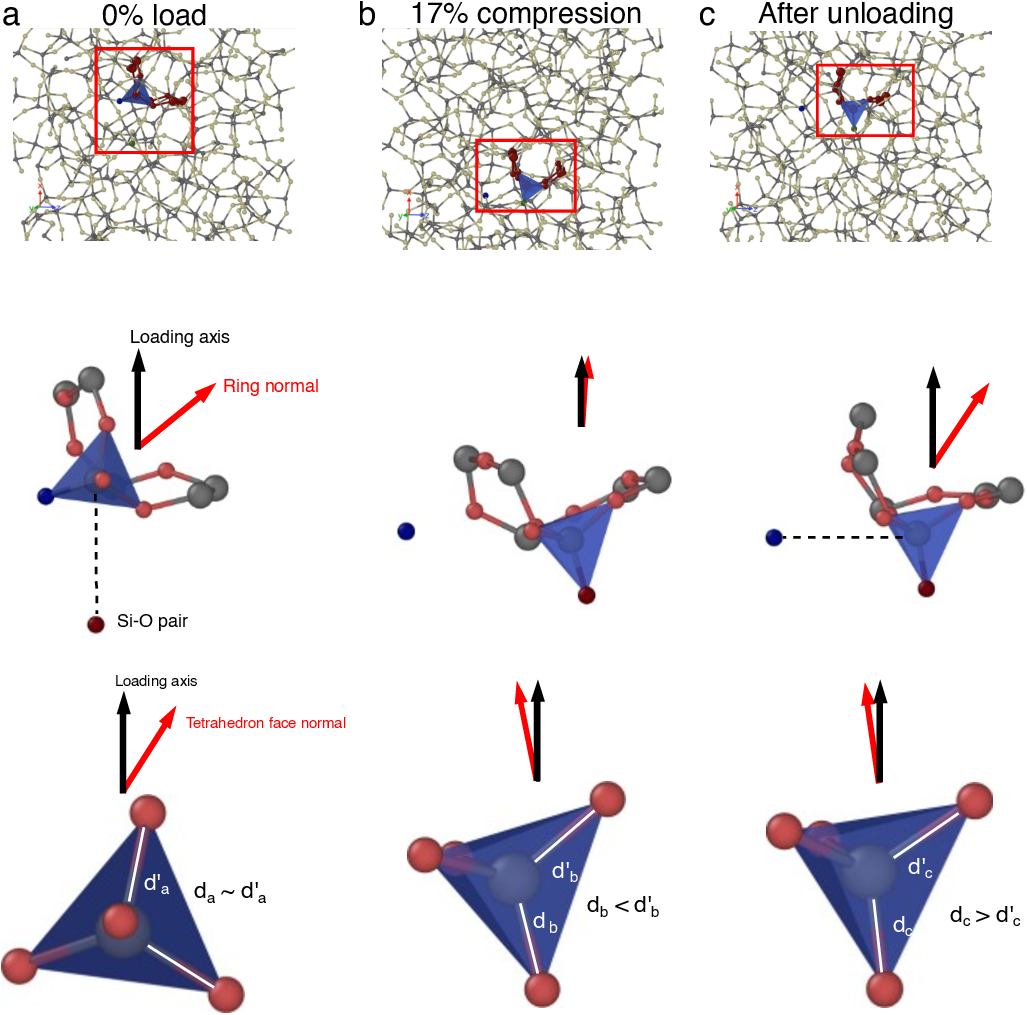}
	\caption{\label{bondswhitching}\footnotesize 
		Structural changes and mechanisms causing transient and persistent anisotropy during the MD simulation of uniaxial compression and unloading at RT. Always the same tetrahedron and ring are shown within a thin slice of the system. Changing positions are due to the homogeneous deformation. 
		Details from configurations at (a) 0\% strain, (b) at 17\% compressive strain, and (c) after unloading from 17\%  compressive strain.
		First row: local detail of the silica glass network structure (grey: Si atoms, light yellow: O atoms), red Si and O atoms are used to highlight a 6-membered ring, a SiO$_4$ tetrahedron from this ring is highlighted in blue, the green O atom is initially part of another tetrahedron. 
		During compression to 17\% strain, the central Si atom of the depicted tetrahedron changes its bonding topology by switching one bond to the green O atom.
		Second row: zoom in on the individual ring and tetrahedron, showing the reorientation of the ring normal, and highlighting the bond loss in the direction orthogonal to the compressive axis and the formation of a new bond along the loading axis, that does not change back to the original configuration. The entire process is shown in movie M1. 
		Third row: zoom in on the tetrahedron, showing the change of the tetrahedra face normal orientation and comparing the lengths of Si--O bonds: in the unstrained configuration (a) all bonds have similar lengths; in (b) the newly formed  bond parallel to the loading axis is shorter than bonds that are more orthogonal to the loading axis; in (c) this bond is longer than bonds orthogonal to the loading axis.}
\end{figure}

\clearpage

\subsection{Direct observation of mechanisms leading to anisotropy}

The mechanisms and structural changes causing anisotropy can not only be inferred from analysis averaging over the entire sample, Figures~\ref{Stressanisofact} and \ref{bondstatus}, but can also be directly observed in the simulations.
Figure~\ref{bondswhitching} and movie M1 show the process in detail, where during compression to 17\% strain, the central Si atom of the shown tetrahedron switches a bond to a nearby O atom.
This leads to a bond loss in the orthogonal direction to the compression axis and a new bond along the loading axis. During unloading, the newly formed bond does not switch back to the original configuration, but the tetrahedron rotates to adjust for the increase in sample length during unloading.
The middle and lower row of Fig.~\ref{bondswhitching} show in addition the change of orientation of the ring and tetrahedra normals as well as the relative bond length of differently oriented Si--O bonds of the tetrahedron.
This provides direct evidence for our interpretation of Figures~\ref{Stressanisofact}(d-e) and \ref{bondstatus}(c, d).

\subsection{Anisotropy through high-temperature processing}

\begin{figure}[!ht]
	\centering
	\includegraphics[width=0.8\textwidth]{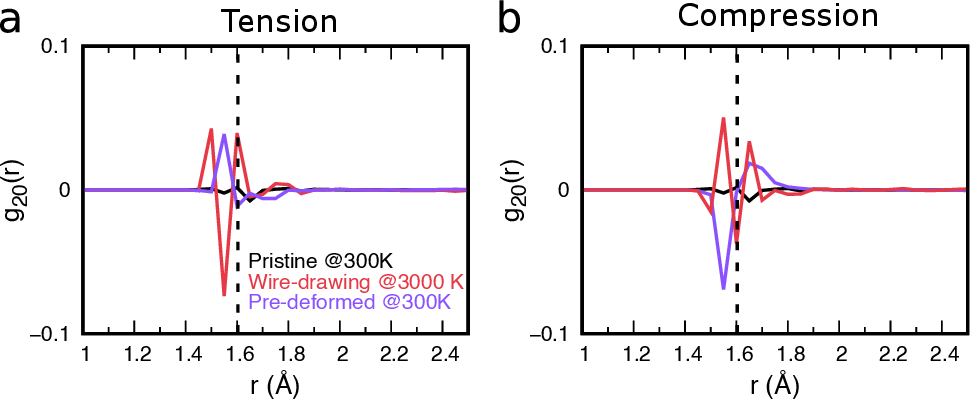}
	\caption{\label{g20_high_temp_pristinet} Projection of the Si-O pair distribution function onto the spherical harmonic $Y_{20}$ function comparing glasses with anisotropy induced through wire-drawing-like process at $T_g$ and by mechanical pre-deformation at RT in tension (a) and compression (b). Black dashed lines indicate the average Si--O bond length in the pristine glass. Solid black: unstrained samples; purple: deformed to 17\% strain and unloaded at RT, compare Fig.~\ref{Stressanisofact}(c, d);
		red: pre-deformed to 17\% strain at $T_g$ and subsequently quenched and relaxed.
		See the discussion in the main text and of Fig. \ref{Stressanisofact} for the interpretation of $g_{20}(r)$.}
\end{figure}

In many industrial processes like fiber- or sheet-drawing, deformation takes place at temperatures above $T_g$. In order to mimic such experimental conditions, we have prepared anisotropic samples by applying the deformation at high-temperatures (HT), i.e., near $T_g$, similar to the procedure in Ref.~\cite{Bidault2016}, see the methods section for more details.
The stress-strain curves, the transient and persistent anisotropy for different deformation temperatures are shown in Fig. S4 in the supplementary material.
While the transient anisotropy is like the stress response systematically lower with increasing temperature, the magnitude of persistent anisotropy after quenching from 3000 K to 300 K and relaxing can -- for tension -- be similar to the one at 300K. However, for compressive loading, the persistent anisotropy does not reach the same high values as for 300 and 600 K.
The anisotropy of these samples as characterized by $g_{20}(r)$ is shown in Fig.~\ref{g20_high_temp_pristinet}(a and b) together with the samples pre-deformed at RT. 
This plot clearly shows that the persistent anisotropy of glasses prepared through different routes has different structural origins; i.e., while the situation is more complex for the high-T deformed glasses than the RT deformed ones, 
the majority of short Si-O bonds are oriented after tensile predeformation at 3000 K orthogonal to the deformation axis, and parallel for compression. 
The situation is exactly the opposite for RT-predeformed glasses. 

 The measured transient anisotropy of these samples are significantly lower than 
 for the glasses deformed at lower temperatures.
 This is attributed to the viscous nature of the high-temperature deformation that involves much more bond breaking than during the mostly plastic deformation at RT, see Fig.~S9 in the supplementary material. 
 This prevents the formation of larger, anisotropic structures.
 This is also evident from Fig.~S10 of the supplementary material, which shows in contrast to the situation at 300 K, at 3000 K no preferential orientation of Si-O bonds during loading is discernible.  
 It is therefore not surprising that the quenched and relaxed samples have also a generally low value of the persistent anisotropy factor $\alpha_\text{p}$, see Fig. S4 in the supplementary material.
 However, for tensile deformation, the HT-quench method seems to yield slightly higher anisotropy than in compression and also than the tensile plastic deformation at RT, see Fig. S4 in the supplementary material. 
 This method -- akin to wire-drawing -- thus yielded the highest persistent anisotropy in tension. 
 This method of inducing anisotropy is expected to be even more interesting in glasses with partially-polymerized structures like in meta-silicate and meta-phosphate glasses as discussed in section \ref{structural_origin}.
 The higher persistent anisotropy after compression at RT is due to the anomalous compressibility of pure silica glass, which allows for high plastic strains. 
 Introducing anisotropy by compression is therefore best performed at RT.
 
\newpage

\section{Conclusion}
Using novel structural analysis methods, we demonstrate that during MD simulations of uniaxial tension and compression at RT, the silica tetrahedra and the Si--O rings reorient themselves with respect to the loading axis, leading to transient anisotropy. 
During plastic deformation, only relatively few atoms change their bonding partners. 
The few bonds that broke during deformation were, however, preferentially oriented with respect to the loading axis: parallel for tension and orthogonal for compression. 
In contrast, the newly formed bonds were oriented orthogonal to the deformation axis for tension and parallel for compression. 
Few of these newly established bonds change during unloading, leading to persistent anisotropy in the stress-free glass. 
Persistent anisotropy is more pronounced after compression than after tension, as larger plastic strains can be reached in the former, while the glass shows limited plasticity and fractures under tension.
Furthermore, it is shown that persistent anisotropy induced by mimicking a fiber-drawing process at $T_g$ is structurally different from persistent anisotropy that results from deformation at RT.

The presented mechanistic insights into the emergence of different types of persistent anisotropy through different processing routes will enable new strategies for tailoring oxide glass properties and for designing new glasses for advanced technological applications.

\section*{Acknowledgements}
Financial support of this work was obtained from the German Science Foundation DFG through its priority program SPP 1594 ”Topological Engineering of Ultra-strong Glasses”. A.A, E.B and L.W. acknowledge funding from the European Research Council (ERC) under the European Union’s Horizon 2020 research and innovation programme (grant agreements No 725483 (A.A. and E.B.) and 681652 (L.W.)). J. G. acknowledge funding from the French ANR, grant ANR-21-CE08-0001. A.A thanks Dr.-Ing. Duancheng Ma for fruitful discussions. S.G. thanks Prof. Johannes Roth, Prof. James Kermode, Dr. Thomas Zeiser, and Dr. Faisal Shahzad for their help in the implementation and parallelization of silica force-fields in the IMD software package. The authors gratefully acknowledge the computer resources provided by the Erlangen Regional Computing Center (RRZE).

\bibliographystyle{elsarticle-num} 
\bibliography{references}

\end{document}


\begin{frontmatter}

		\title{\textbf{Supplementary Materials}\\
			The Origin of Deformation Induced Topological Anisotropy\\ in Silica Glass}
		
\author[fau,lem3,aau]{Sudheer Ganisetti}
\author[fau,max,uds]{Achraf Atila}
\author[lem3]{Julien Gu\'{e}nol\'e}
\author[tubaf]{Aruna Prakash}
\author[hhu]{J\"{u}rgen Horbach}
\author[fsu]{Lothar Wondraczek}
\author[fau,max]{Erik Bitzek\corref{cor1}}

\cortext[cor1]{e.bitzek@mpie.de}

\affiliation[fau]{organization={Department of Materials Science and Engineering, Institute I,Friedrich-Alexander-Universität Erlangen-Nürnberg (FAU)},
	city={Erlangen},
	postcode={91058},
	country={Germany}}
 
\affiliation[lem3]{organization={Université de Lorraine, CNRS, Arts et Métiers, LEM3},
	city={Metz},
	postcode={57070},
	country={France}}
  
\affiliation[aau]{organization={Department of Chemistry and Bioscience, Aalborg University},
	city={Aalborg},
	postcode={9220},
	country={Denmark}}

\affiliation[max]{organization={Computational Materials Design, Max-Planck-Institut für Eisenforschung},
	city={Düsseldorf},
	postcode={40237},
	country={Germany}}

\affiliation[uds]{organization={Department of Material Science and Engineering, Saarland University},
	city={Saarbrücken},
	postcode={66123},
	country={Germany}}
 
\affiliation[tubaf]{organization={Micromechanical Materials Modelling (MiMM), Institute of Mechanics and Fluid Dynamics, Technische Universität Bergakademie Freiberg (TUBAF)},
	city={Freiberg},
	postcode={09599},
	country={Germany}}

\affiliation[hhu]{organization={Institut für Theoretische Physik II, Heinrich-Heine-Universität Düsseldorf},
	city={Düsseldorf},
	postcode={40225},
	country={Germany}}

\affiliation[fsu]{organization={Otto Schott Institute of Materials Research, Friedrich Schiller University Jena},
	city={Jena},
	postcode={07743},
	country={Germany}}
				
	\end{frontmatter}

\textbf{This PDF file includes:}

Supplementary Text

Figs. S1 to S10

Tab. S1

References cited in the supplementary text.\\

\textbf{Other Supplementary Materials for this manuscript include the following: }

Movie M1.mpg showing a bond-switching process leading to structural re-orientation. 

\newpage

\section*{Effect of simulation parameters}

\subsection*{Effect of the interatomic potential}

\begin{figure}[h]
	\centering
	\includegraphics[width=0.8\textwidth]{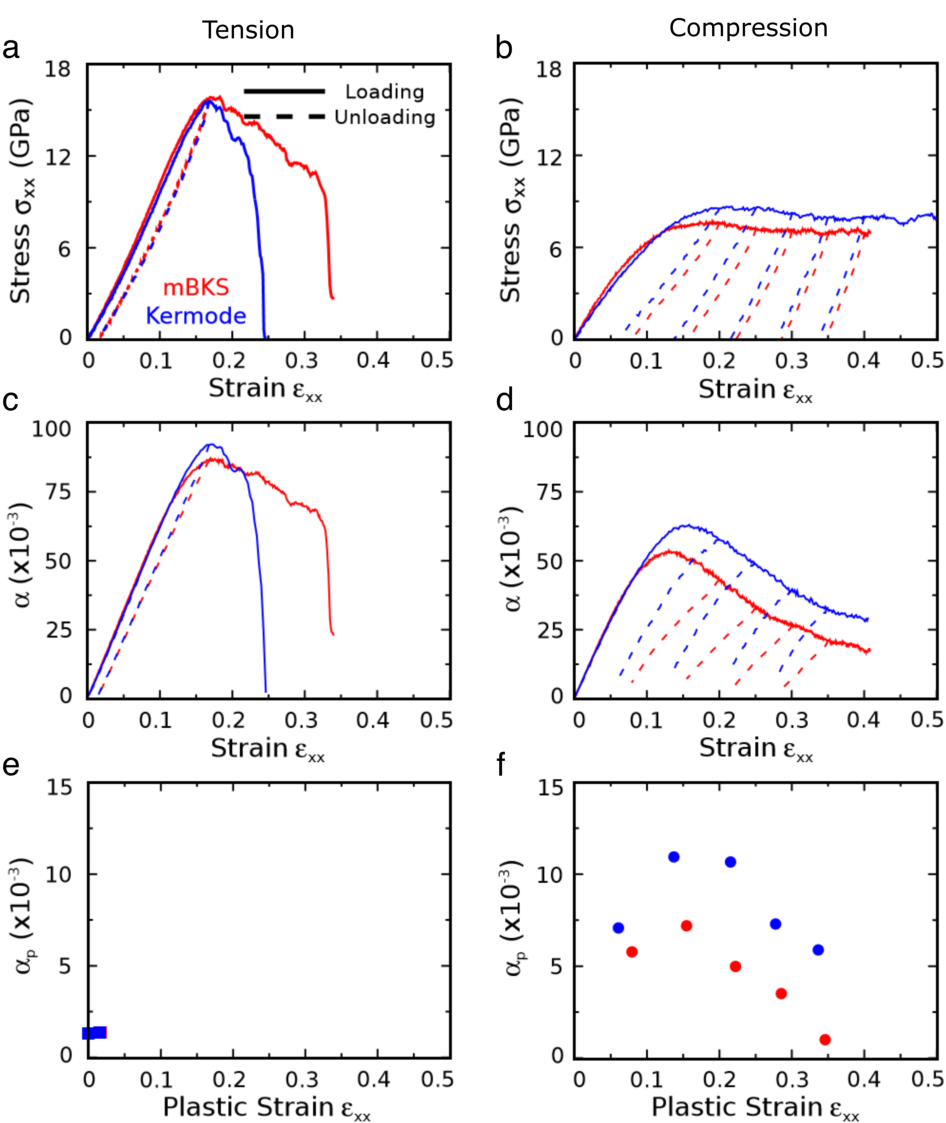}
	\caption{\label{potential_effect}Top row shows the stress over strain, middle row shows the anisotropy during load, and bottom row shows the persistent anisotropy as a function of the plastic strain for (first column) tension and (second column) compression with two different potentials, namely Kermode (red) and BKS (blue).}
\end{figure}

Figure~\ref{potential_effect} shows the stress-strain curves, $\alpha$ as function of strain and the persistent anisotropy index $\alpha_p$ for samples simulated with the Kermode~{\cite{Kermode2010}} and the mBKS~{\cite{vanBeest1990}} potentials.
The elastic response of both potentials is very similar, however the mBKS potential shows higher ductility in tension and a somewhat lower flow stress in compression than the Kermode potential.
The transient anisotropy $\alpha$ under load is lower for the mBKS than the Kermode potential.
The same trend is visible for $\alpha_p$.

\subsection*{Effect of the strain rate}

\begin{figure}[h]
	\centering
	\includegraphics[width=0.8\textwidth]{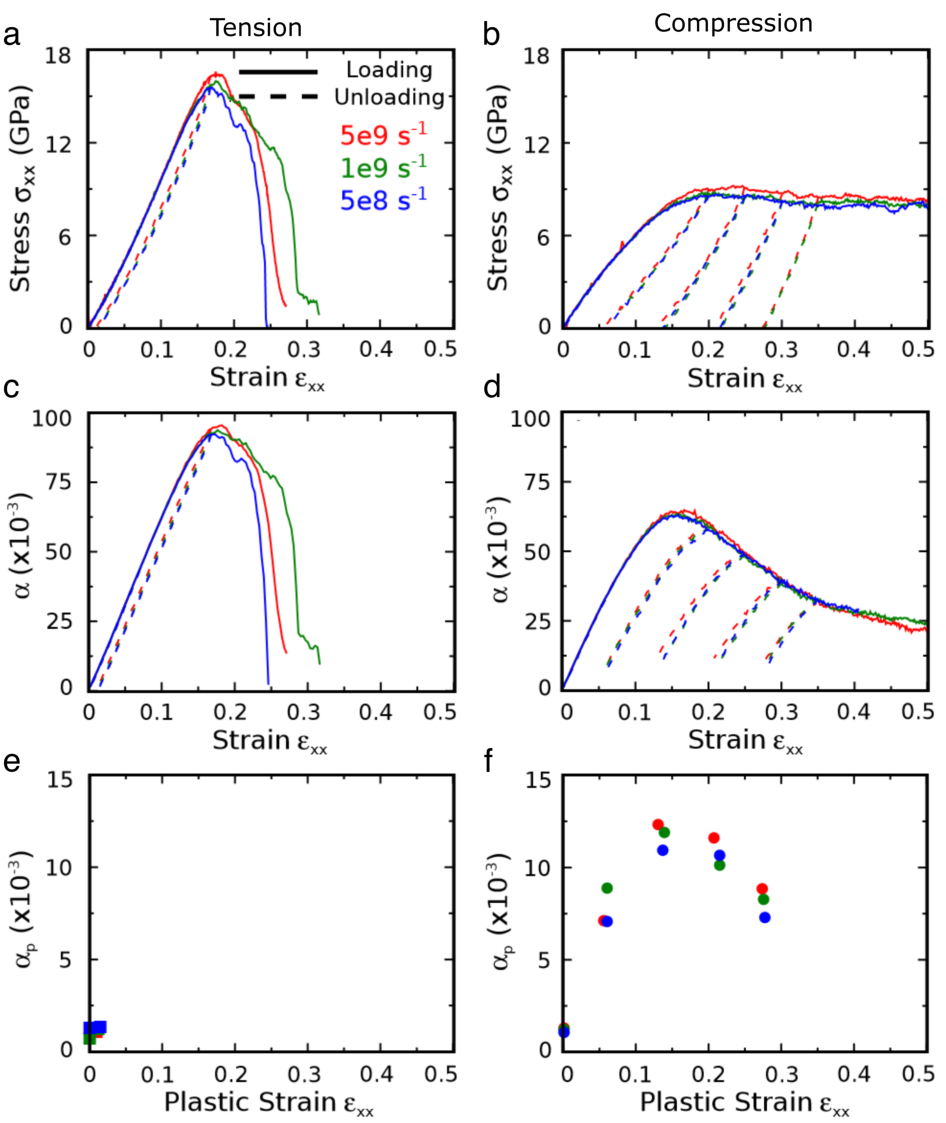}
	\caption{\label{strain_rate_effect}
		The top row shows the stress-strain curves, the middle row shows the anisotropy during the loading, and the bottom row shows the persistent anisotropy as a function of the plastic strain after unloading for (first column) tension and (second column) compression using three different strain rates of 5$\times$10$^8$ s$^{-1}$ (red), 10$^9$ s$^{-1}$ (blue) and 5$\times$10$^9$ s$^{-1}$ (green).}
\end{figure}

Figure~\ref{strain_rate_effect} shows the stress-strain curves, $\alpha$ as function of strain and the persistent anisotropy index, $\alpha_p$ for the same sample deformed at different strain rates.
As can be seen, the ultimate strength is increased with increasing strain rate, which is typical for thermally activated events.
The (transient) anisotropy parameter as function of strain is nearly not affected by the deformation rate,
The persistent anisotropy index $\alpha_p$ shows some variation, however without clear trend as function of strain rate.

\subsection*{Effect of the cooling rate}

\begin{figure}[h]
	\centering
	\includegraphics[width=0.8\textwidth]{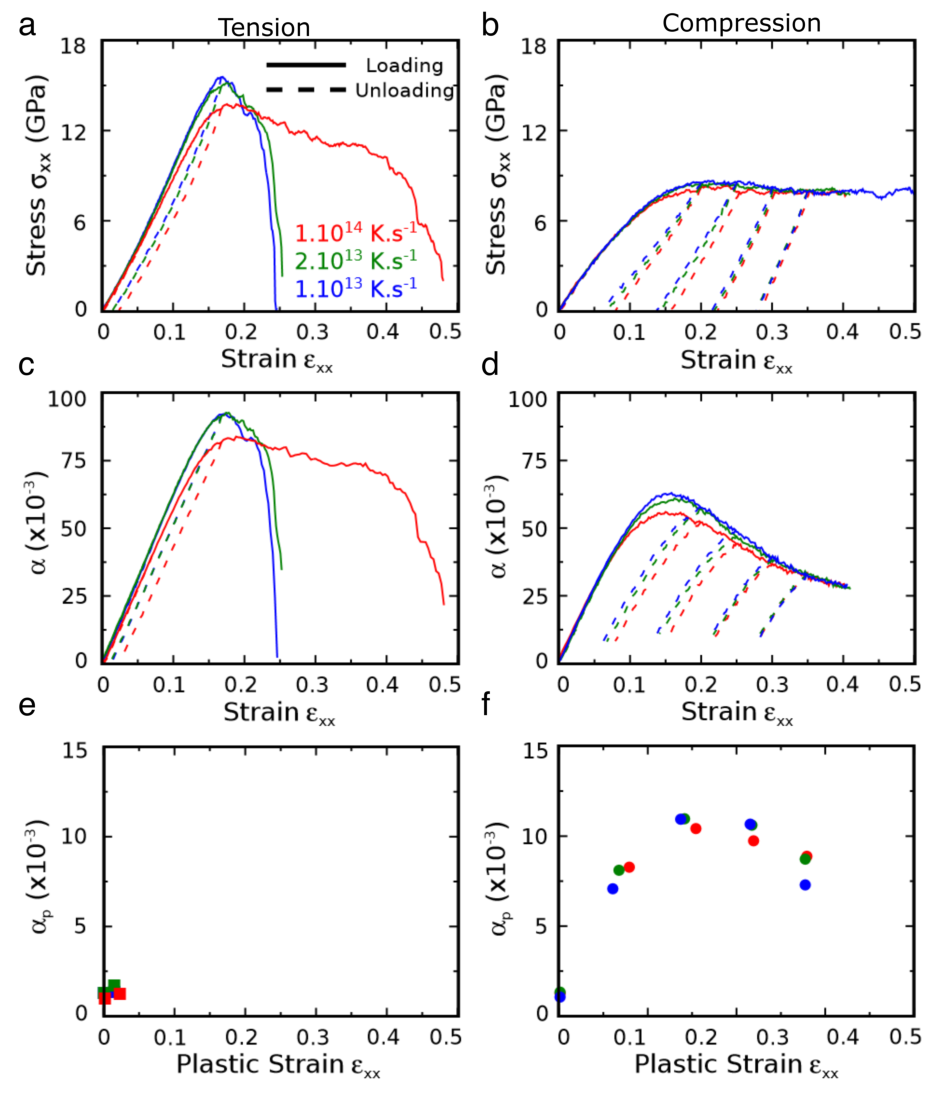}
	\caption{\label{cooling_rate_effect}Top row shows the stress-strain curves, middle row shows the anisotropy during the loading, and bottom row shows the persistent anisotropy as a function of the plastic strain after unloading for (first column) tension and (second column) compression for the samples prepared with three different cooling rates of 10$^{13}$ K s$^{-1}$ (red), 2$\times$10$^{13}$ K s$^{-1}$ (blue) and 10$^{14}$ K s$^{-1}$ (green).}
\end{figure}

The mechanical behavior of glasses depends on the cooling rate used for the preparation of the glass. Figure~\ref{cooling_rate_effect}(a) and Figure~\ref{cooling_rate_effect}(b) show the tensile and compressive stress-strain curves of silica glass prepared using different cooling rates. 
The strength of the glass prepared using a higher cooling rate is generally lower than for slowly cooled glasses.
Conversely, the fracture strain in tension increases with cooling rate.
This is a general observation and attributed to the fact that rapidly cooled glasses are less relaxed and have more defective (i.e., over-coordinated) atoms than the slowly cooled glasses and show a lower stiffness compared to the slowly quenched glasses~{\cite{Zhang2020}}. 
As the plastic strain after unloading is different for the differently cooled samples, 
the persistent anisotropy indices $\alpha_p$ are somewhat different, but follow roughly the same trend as a function of plastic strain.

\subsection*{Effect of deformation temperature}

\begin{figure}[h]
	\centering
	\includegraphics[width=0.8\textwidth]{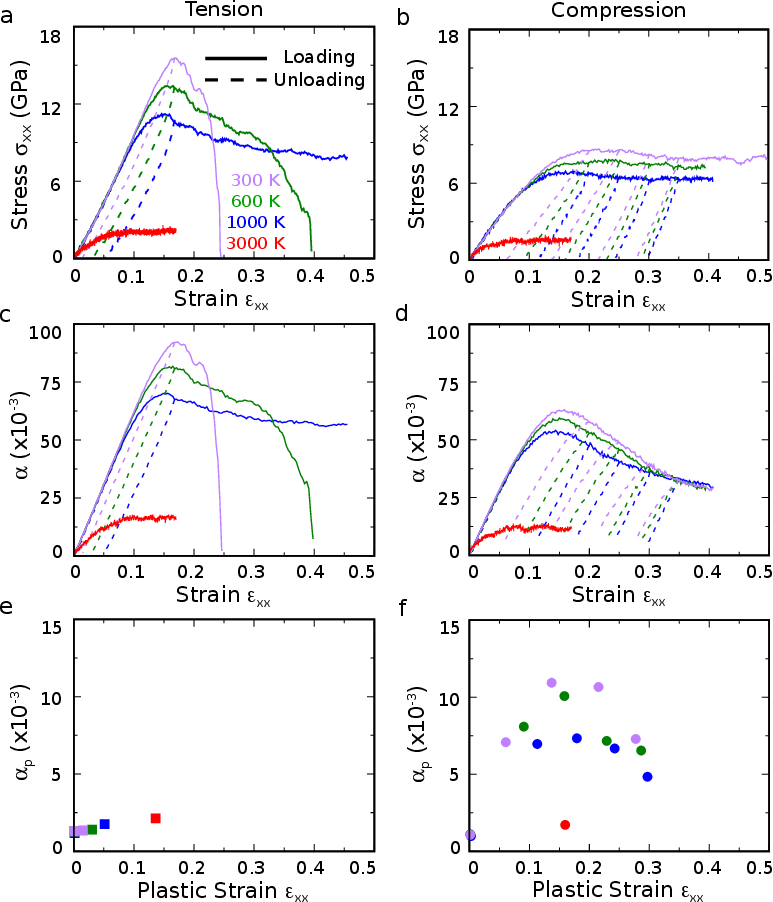}
	\caption{\label{temperature_effect}The top row shows the stress-strain curves, the middle row shows the anisotropy during the loading, and the bottom row shows the persistent anisotropy as a function of the plastic strain after unloading for (first column) tension and (second column) compression at temperatures of $300$ K (purple), $600$ K (green), $1000$ K (blue). The red data is from deformation at $3000$ K mimicking the wire-drawing process, i.e., instead of unloading at the same temperature, the glass is quenched to 300 K and the stress is relaxed.}
\end{figure}

The stress-strain curves, $\alpha$ as function of strain and the persistent anisotropy index, $\alpha_p$ for the same sample deformed at different temperatures, are shown in Figure~\ref{temperature_effect}. 
As expected for glasses, the yield strength is reduced and the plasticity increased with increasing temperature. 
The reduced flow stresses are correlated with reduced $\alpha$ values during straining.
At higher temperatures, significant plastic deformation also becomes possible in tension.
However, $\alpha_p$ from the tensile samples at 1000K shows still very low values, although a strain of about 5\% could be reached. 
For comparable values of plastic strain, $\alpha_p$ in compression shows some variation with deformation temperature.
In particular for large plastic strains $\alpha_p$ seem to decrease with increasing deformation temperature.
It has, however to be noted, that the temperatures here are well below the glass transition temperature of $T_g=3000$K.

\subsection*{Conclusions}
While the magnitudes of transient and persistent anisotropy as determined by the fabric tensor do depend on
the potential, the cooling rate used to create the sample, the deformation rate and temperature, the overall behavior is comparable: all samples deformed in compression show persistent anisotropy, which for plastic strains up to about 15\% increases with plastic strain and decreases with larger plastic strain. 
The largest differences in transient anisotropy are seen for different deformation temperatures (about 1/3 reduction from 300K to 1000K in tension) and for different potentials. 
The largest differences in persistent anisotropy are caused by the use of different potentials.
No consistent trend in the influence of temperature, strain rate and cooling rate on $\alpha_p(\epsilon)$ can be established, and for similar plastic strains variations of $\alpha_p$ of about 10-20\% are observed.

\clearpage
\section*{Stereographic projection}

\begin{figure}[h]
	\centering
	\includegraphics[width=0.8\textwidth]{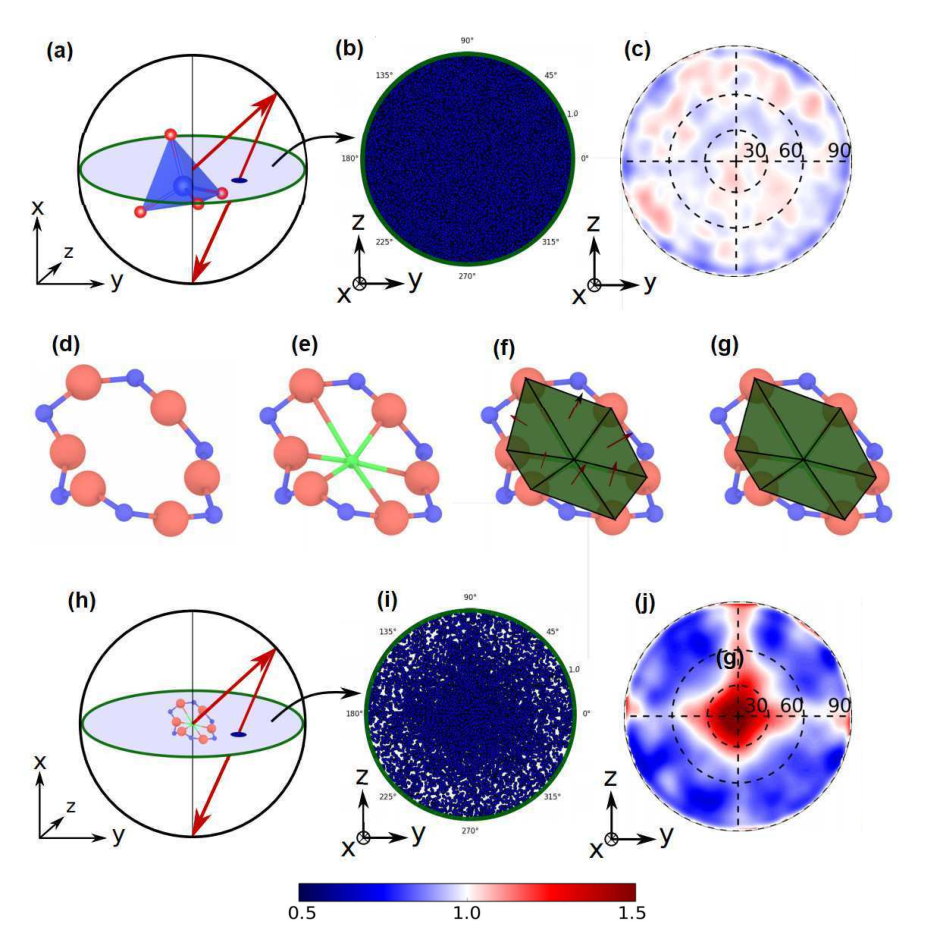}
	\caption{\label{steriographicprojmet}
		Illustration of the method to obtain stereographic projection of the normals of tetrahedra faces rings. 
		(a) Identification of tetrahedra and construction of the stereographic projection for one of its face normals
		on the $yz$ projection plane; (b) resulting scatter plot on the projection plane for all tetrahedra face normals;
		(c) point density of the stereographic projection of tetrahedra face normals normalized by the point density
		for a random distribution of orientations. No preferred orientations are discernable.
		(d) Identified ring; (e) determine center of mass (green), (f) ring discretized into triangles with two neighboring Si atoms and center of mass and normals of each triangle; (g) the ring normal is defined as the average of all triangle normals; 
		(h) stereographic projection of the ring normal on the $yz$ plane; (i)  resulting scatter plot on the projection plane for all ring normals (i) normalized stereographic projection of ring normals showing a clear preferred orientation along the pole. (b, c, i and i) are plotted for a  17\% compressed sample of 72000 atoms.}
\end{figure}

The stereographic projection is a way of mapping that projects vectors pointing at the surface of a sphere onto a plane. 
This method is extensively used within the crystallography community to analyze the microstructure of materials. 
To the best of our knowledge, this is the first time this method has been applied to oxide glasses. 
In the present study, the orientation of tetrahedra face normals, ring normals, and bonds were investigated using stereographic projections. 
A complete theoretical foundation of this method can be found in textbooks such as Refs.~{\cite{kelly2020crystallography, whittaker1984stereographic}}. 
The steps that we used to adapt this method to silica glass are detailed below, and the schematic pictures in Figure~\ref{steriographicprojmet} will assist in following the steps.

\begin{enumerate}
\item Normalization of the vector.
\item Construction of a unit sphere with the same origin as the vector.
\item Construction of a cutting plane dividing the unit sphere in two equal hemispheres. This is the projection plane. The intersection points of the projection plane normal with the upper and lower hemisphere are called poles.
\item Construct a line from the intersection of the vector and the upper hemisphere to the pole of the lower hemisphere. 
\item Mark the point of intersection of this line with the projection plane.
\item Repeat for all vectors.
\end{enumerate}
When a large number of vectors with different orientations are projected, the total area of the projection map will be full of points, e.g., Figure~\ref{steriographicprojmet}(b). 
Therefore, the local point density is color coded.
As the stereographic projection is not an equal-area projection,
the point density needs to be normalized by the point density expected for a random distribution of orientations.
This density was calculated numerically using random vectors. 
The normalized density is color-coded and each point on the projection map can be addressed by zenith $\theta$ and azimuthal $\phi$ angles.

\clearpage

\section*{Glass structure}
\begin{figure}[h]
	\centering
	\includegraphics[width=0.8\textwidth]{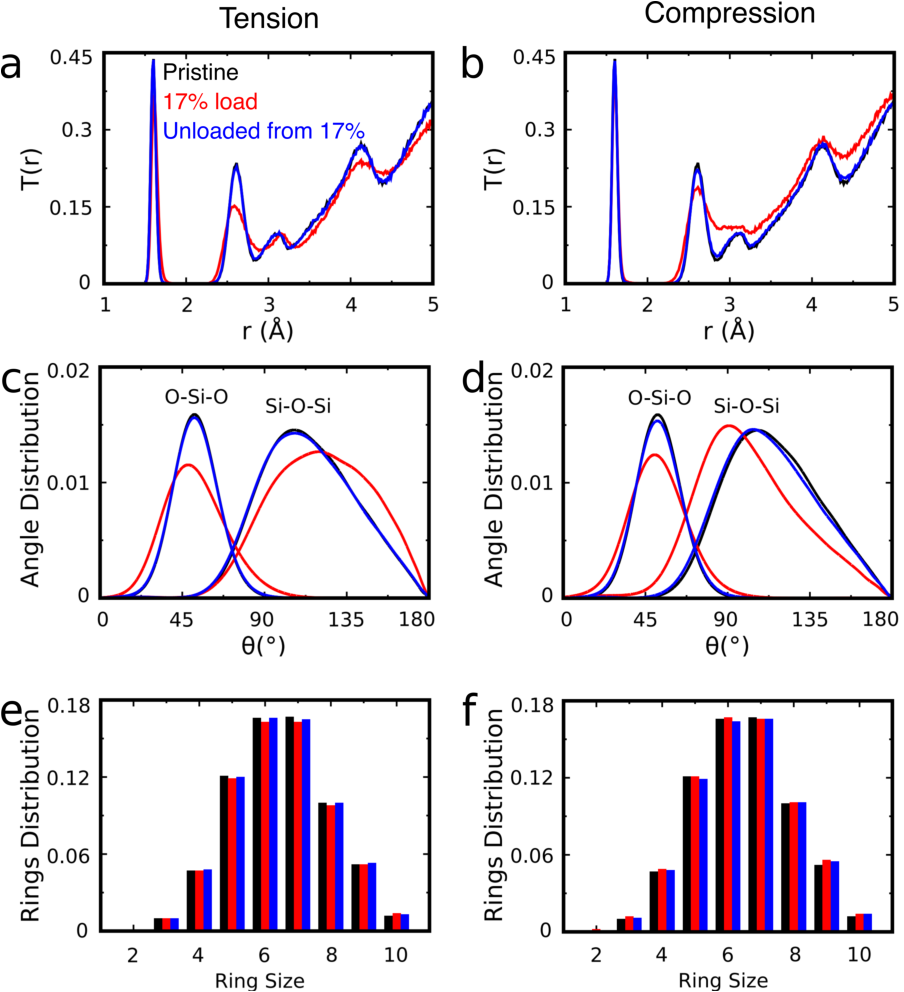}
	\caption{\label{Structure}Top: radial distribution function of the pristine, loaded, and unloaded samples in both tension (a) and compression (b). 
		Middle: bond angle distribution (smoothed data points) for O--Si--O and S--O--Si bonds in the pristine, loaded, and unloaded glasses in tension (c) and compression (d). 
		Bottom: ring size distribution of pristine, loaded, and unloaded samples in  tension (e) and compression (f).}
\end{figure}

The total radial distribution, $T(r)$, of the pristine silica glass, is shown in Figure~\ref{Structure}, and is in good agreement with the data available in the literature from both experiments~{\cite{Greaves2007}} and simulations~{\cite{Kermode2010}}.
A comparison of  $T(r)$ for the pristine, loaded, and unloaded glasses is provided in Figure~\ref{Structure}(a) and Figure~\ref{Structure}(b).
It shows that the position of the first peak, which is centred around 1.603 \AA\ and corresponds to the Si--O bond, is slightly affected. 
The positions of the peaks that correspond to O--O and Si--Si bonds, found at larger distances, are more affected by the loading while they recover their initial position in the unloaded samples. 
The bond angle distributions of O--Si--O and Si--O--Si bonds for the pristine, loaded, and unloaded glasses are shown in Figure~\ref{Structure}(c) and Figure~\ref{Structure}(d). 
The mean O--Si--O and Si--O--Si bond angles in the pristine glass are 109.2\textdegree{} and 143\textdegree{}, respectively. Both angles show a significant change in the loaded samples, and they recover their initial state in the unloaded samples. 

In silica glass, the SiO$_4$ tetrahedra are connected to form a network made of rings of different sizes. 
The RINGS code~{\cite{LeRoux2010}} was used to calculate the ring size distribution, where the ring size is defined as the number of Si atoms that belong to a given ring. 
Figure.~\ref{Structure}(e) and Figure~\ref{Structure}(f) show the ring size distribution for pristine, loaded, and unloaded glasses. 
The 6 and 7 membered-rings (MR) were more frequent than rings of other sizes. 
Minor changes in the ring distribution were found when comparing the pristine, loaded, and unloaded samples. 
The number of 3, 4 and 9 MR was almost constant; 5, 6, 7 and 8 MR slightly decreased; the population of 9 and 10 MR slightly increased in the tensile loaded sample. 
However, the difference in the population of any numbered ring between the pristine and unloaded samples from 17\% tensile strain was negligible.
The population of all rings, except 5 and 7 MR, were increased in the sample loaded in compression.
The population of the 5, 6 and 7 MR were lower, and 8, 9, and 10 MR were higher in the unloaded sample compared to the pristine sample. 

\section*{Characterization of anisotropy by fabric tensor analysis}
\begin{figure}[h]
	\centering
	\includegraphics[width=0.8\textwidth]{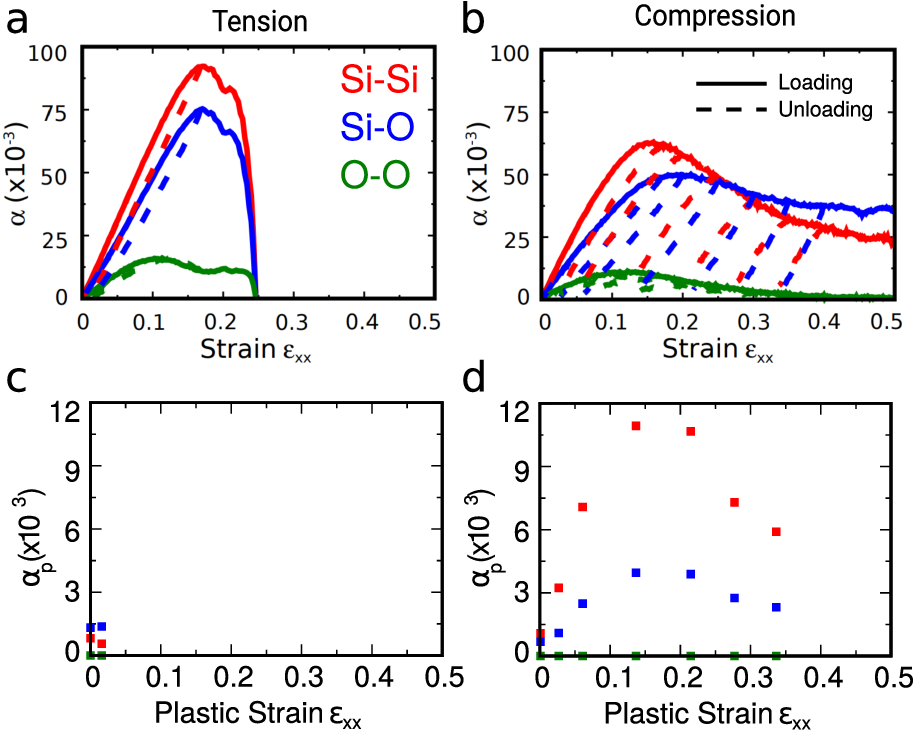}
	\caption{\label{Aniso}(a and b) Evolution of transient anisotropy as captured by the anisotropy factor $\alpha$ during tension and compression. Red, blue, and green curves show $\alpha$ for Si-Si, Si-O, and O-O pairs, respectively. Solid lines represent the loading part, dashed lines show the evolution during unloading from different strains. (c and d) Persistent anisotropy factor $\alpha_p$ after unloading as a function of the plastic strain of the stress-free samples initially loaded in tension or compression, using the same color scheme as in (c and d). Deformation temperature: 300 K, strain rate: $5\times10^{-8}$ s$^{-1}$.}
\end{figure}

The anisotropy index corresponding to Si--Si, Si--O and O--O atom pairs was computed during loading and unloading in tension and compression. These are referred to as $\alpha$(Si-Si), $\alpha$(Si-O), and $\alpha$(O-O) and are shown in Figure~\ref{Aniso}(a and b). 
In both deformation modes, $\alpha$(O-O) is much lower than $\alpha$(Si-Si) and $\alpha$(Si-O), therefore, 
$\alpha$(O-O) will not be specifically addressed.
The anisotropy indices $\alpha$(Si-Si) and $\alpha$(Si-O) follow the same trend as the stress-strain curve in tension but not in compression (compare to Figure~1a,b in the main paper).
During compression, both $\alpha$(Si-Si) and $\alpha$(Si-O) reach a maximum and decrease, 
with the rate of the decrease being higher for $\alpha$(Si-Si) than for $\alpha$(Si-O).
In compression, a finite plastic strain is recorded after unloading from strain values larger than 5\%.
The index of persistent anisotropy for stress-free, unloaded structures is in the following, denoted by $\alpha_p$. 
It is shown in Figure~\ref{Aniso}(c and d) as a function of remaining plastic strain.
The negligible plastic strain achieved in tension is correlated with values of $\alpha_p$ that are comparable to 
the $\alpha$ values of the pristine glass. 
However, unloading from higher strain in compression results in glasses with higher plastic strains and values of $\alpha_p$(Si-Si) and $\alpha_p$(Si-O) that are higher than those of the pristine glass. 
The index of persistent anisotropy for  samples unloaded from compression show a maximum at a plastic strain of about 15\%, indicating that deformation in the plastic flow regime does not further contribute to persistent anisotropy, but instead reduces it.

\clearpage

\begin{figure}[h]
	\centering
	\includegraphics[width=1.0\textwidth]{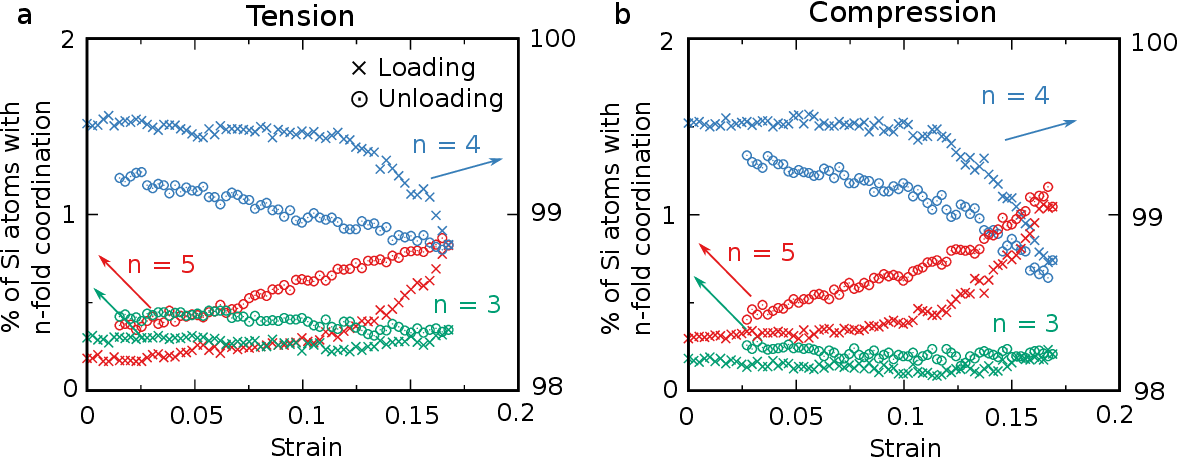}
	\caption{\label{strain_coord} Evolution of 3-, 4-, and 5-fold coordinated Si as a function of strain during deformation at 300 K of pristine silica under (a) tension (b) compression. Strain rate: $5\times10^{-8}$ s$^{-1}$.}
\end{figure}

\begin{figure}[h]
	\centering
	\includegraphics[width=1.0\textwidth]{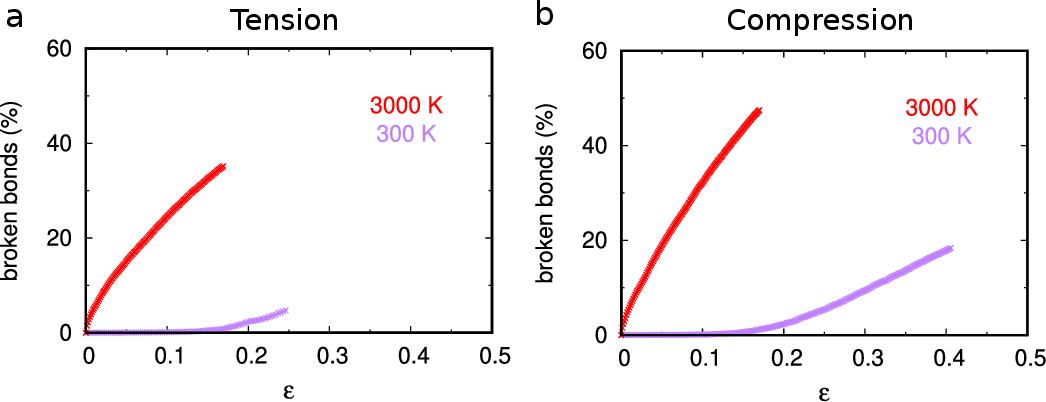}
	\caption{\label{compare_brokenbonds_300K_3000K} Comparison of broken bonds ($\%$) over strain during deformation at 300 K and 3000 K under (a) tension (b) compression. Strain rate: $5\times10^{-8}$ s$^{-1}$.}
\end{figure}

\clearpage

\begin{figure}[h]
	\centering
	\includegraphics[width=0.8\textwidth]{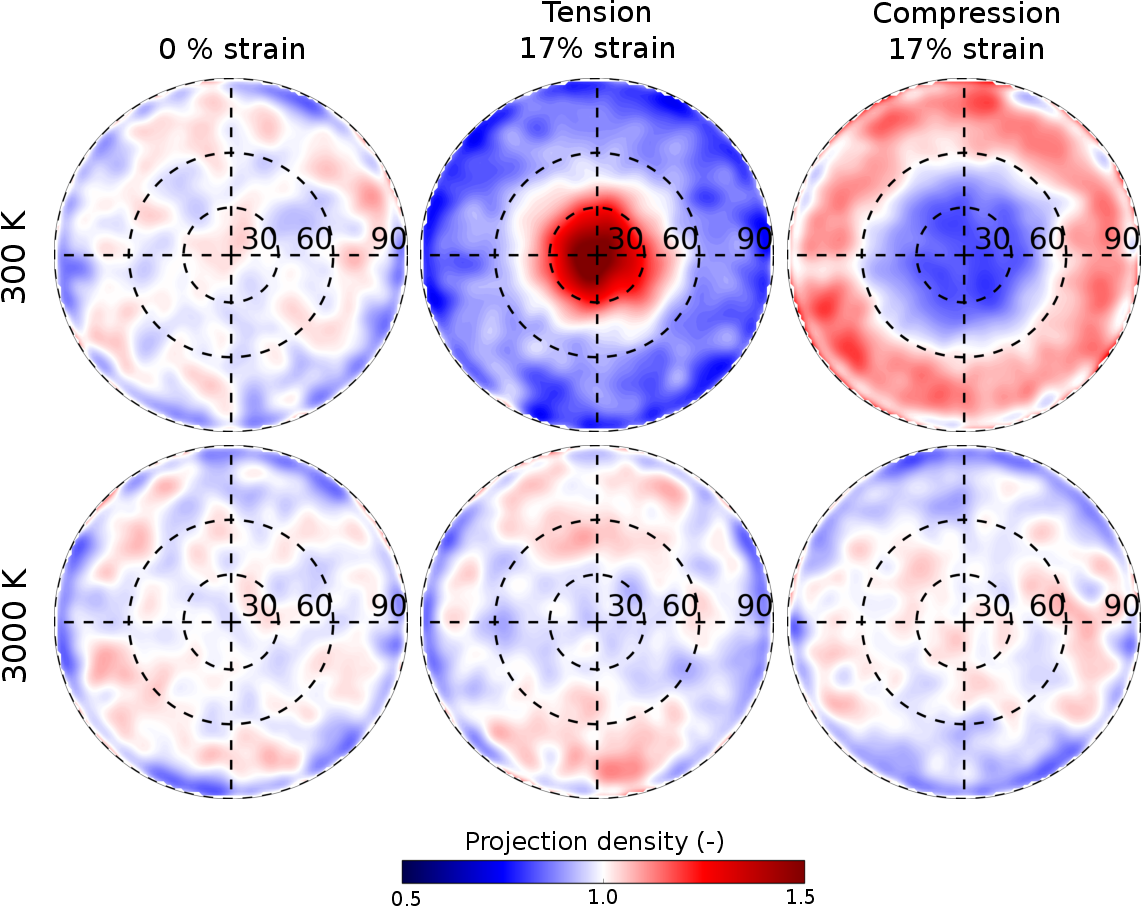}
	\caption{\label{hot_cold_comp_tens_polar_plots} Stereographic projection of Si-O bonds at different temperature and strain states. First row: deformation at $300$ K, second row: deformation at $3000$ K. First column: at $0 \%$ strain, second column: at $17 \%$ tensile strain, third column: at $17 \%$ compressive strain. Strain rate: $5\times10^{-8}$ s$^{-1}$.}
\end{figure}
\clearpage

\begin{table*}[!h]
\centering
\caption{List of uni-axially and hydrostatically compressed samples at $300 K$ with their stress, strain and density histories. All the samples were prepared with Kermode potential by cooling at a rate of 10$^{13}$ K s$^{-1}$; Deformation temperature: $300$ K, strain rate: $5\times10^{-8}$ s$^{-1}$.}
\centering
\renewcommand{\arraystretch}{1.2}
\begin{tabular}{ccccc}
\hline
		sample  & maximum & maximum & density at & persistent \\
            state   & strain  & stress  & maximum    & density \\
                    & (\%)    & ($GPa$) & stress     & (g/cm$^{-3}$)\\
  		\hline
  		pristine & - & - & - & 2.39 \\
		\hline
		& 2 & 1.41   & 2.42  & 2.39 \\
		  & 5 & 3.44   & 2.48  & 2.39 \\
		& 7 & 4.46   & 2.51  & 2.39 \\
		uni-axial & 17 & 8.28  & 2.73  & 2.43 \\
		compression & 20 & 8.53  & 2.78  & 2.48 \\
		& 25 & 8.50 & 2.85 & 2.59 \\
		& 30 & 8.26 & 2.88 & 2.67 \\
		& 35 & 7.90 & 2.90  &2.71 \\
		& 40 & 8.01 & 2.91  &2.65 \\
		\hline
		hydrostatic & 9.5 & 12.7 & 3.23  & 2.47 \\
		 compression  &  &  &  &\\
		\hline
	\end{tabular}
	\renewcommand{\arraystretch}{1}
	\label{density_table}
	\vspace{-5mm}
\end{table*}

\bibliographystyle{elsarticle-num}
\bibliography{references}